\documentclass[preprint,12pt]{elsarticle}

\usepackage{amssymb}
\usepackage{url}
\usepackage{gensymb} 
\usepackage{amsmath,amsfonts}
\usepackage{algorithmic}
\usepackage{algorithm}
\usepackage{array}
\usepackage{textcomp}
\usepackage{stfloats}
\usepackage{url}
\usepackage{verbatim}
\usepackage{graphicx}
\usepackage{multicol}
\usepackage{subcaption} 

\journal{-}

\newcommand{\INDSTATE}[1][1]{\STATE\hspace{#1ex}}

\newcommand{\INDONE}{\INDSTATE[1]}
\newcommand{\INDTWO}{\INDSTATE[3]}

\newcommand{\action}[1]{\textbf{#1}}

\newcommand{\TimeOracle}{\textsf{\small{Time Oracle}}}
\newcommand{\ClientMeter}{\textsf{\normalsize{Client Meter}}}
\newcommand{\Client}{\textsf{\normalsize{Client}}}
\newcommand{\MeteoFrance}{\textsf{\normalsize{Meteo France}}}
\newcommand{\Veolia}{\textsf{\normalsize{ESCO}}}
\newcommand{\DataQualifier}{\textsf{\normalsize{Data Qualifier}}}
\newcommand{\Predictor}{\textsf{\normalsize{Predictor}}}
\newcommand{\Aggregator}{\textsf{\normalsize{Aggregator}}}
\newcommand{\Validator}{\textsf{\normalsize{Validator}}}
\newcommand{\Function}[1]{\textsf{\normalsize{#1}}}

\begin{document}

\begin{frontmatter}



\title{Reliability Analysis of Smart Contract Execution Architectures: A Comparative Simulation Study}


\author{\"{O}nder G\"{u}rcan}

\affiliation{organization={Norwegian Research Center (NORCE)},
            addressline={Universitetsveien 19}, 
            city={Kristiansand},
            postcode={4630}, 
            country={Norway}}

\begin{abstract}
The industrial market continuously needs reliable solutions to secure autonomous systems. 
Especially as these systems become more complex and interconnected, reliable security solutions are becoming increasingly important.
One promising solution to tackle this challenge is using smart contracts
designed to meet contractual conditions, avoid malicious errors, secure exchanges, and minimize the need for reliable intermediaries.
However, smart contracts are immutable.
Moreover, there are different smart contract execution architectures (namely Order-Execute and Execute-Order-Validate) that have different throughputs.
In this study, we developed an evaluation model for assessing the security of reliable smart contract execution.
We then developed a realistic smart contract enabled IoT energy case study.
Finally, we simulate the developed case study to evaluate several smart contract security vulnerabilities reported in the literature.
Our results show that the Execute-Order-Validate architecture is more promising regarding reliability and security.
\end{abstract}


\end{frontmatter}


\section{Introduction}

One of the appealing applications of blockchains are smart contracts:
code that cannot be changed operating on the blockchain to enable, execute automatically, and guarantee a contract between parties who do not trust each other \cite{Abdelhamid2019,Sharma2023}.
The most salient characteristic of smart contracts is \textit{immutability}: no modifications are allowed after deployment. This means that it remains on the blockchain forever and cannot be misplaced or lost, with the record of the history of transactions calling it, which provides greater transparency and safekeeping \cite{Suvitha2021,Sharma2021}.

The blockchain technology has experienced a rapid rise in popularity due to the introduction of Ethereum's smart contracts \cite{Wood2014}, which have enabled the blockchain to become a general-purpose solution with applications in many different areas.
%
Insurance processes can be improved by triggering claims processing when specific events occur, enabling improved supply networks and the development of new applications in business, commerce, and government.

Despite the advances of smart contracts, several concerns undermine their adoption, especially considering the wide range of attacks that occur by exploiting their vulnerabilities \cite{Yulianto2023}.
One of the most consequent smart contract attacks occurred in 2016, where a Decentralized Autonomous Organization (DAO)\footnote{DAO attack, \url{https://siliconangle.com/2016/06/17/ethereum-dao-attacked-over-55-million-of-ether-cryptocurrency-stolen/}} lost around 2 Million Ether (the equivalent of USD 55 million).
%
Therefore, following the most effective methods for implementing such applications is essential.

Due to their immutable nature, implementation, and testing of smart contracts are not trivial and should be given due consideration for avoiding vulnerabilities related to it \cite{Roussille2023}.
%
In order to achieve this goal, the literature has proposed various Verification, Validation and Testing (VV\&T) solutions for different purposes (for a review, see \cite{Benabbou2021}).
%

Based on this observation, the objective of this paper is to analyse the \textit{operational security and reliability} of smart contracts from the angle of the smart contract execution architectures.
In the world of smart contracts, there are two main execution architectures: \textit{execute-order-validate} and \textit{order-execute}. 
While former first executes a smart contract call by simulating the state change it generates and then includes it in the blockchain if the generated state change is valid, the latter executes a smart contract call after having included it in the blockchain.

Both have their advantages and disadvantages. 
Even though many works are studying the security of the \textit{order-execute} architecture blockchains, there are relatively few works for the \textit{execute-order-validate} architecture ones \cite{Ruan2020,Androulaki2018}.
Moreover, to the best of our knowledge, there is no study comparing them with respect to the same security vulnerabilities.

To this end, we evaluate various security vulnerabilities on a realistic Internet of Things (IoT) use case using smart contracts \cite{Gurcan2018SC}. 
The outcome of this study can be helpful when choosing the right platform to develop a smart contract-based system. 
The contributions of this paper are as follows:

\begin{itemize}
\item We developed a generic simulation model that mimics the {order-execute} and the {execute-order-validate} smart contract execution architectures.
\item We developed a realistic blockchain-enabled IoT use case by extending the developed generic simulation model.
\item Using the developed use case, we empirically analyzed both architectures against various security vulnerabilities covering the three testing levels (i.e., unit, integration and system levels).
\end{itemize}

The remainder of this paper is organized as follows:
Section \ref{sec:Background} provides the background related to blockchain and smart contracts.
Section \ref{sec:Smart-Contract-Execution-Architectures} presents the execution flow of smart contracts considering the available blockchain architectures.
%
In Section \ref{sec:Use-Case} presents a comprehensive model of a smart contract-enabled industrial prototype of trusted energy performance contracts.
In Section \ref{sec:Simulations}, various simulation scenarios are modeled for the security analysis, and Section \ref{sec:Results} presents the results of these simulations.
%
%
Finally, Section \ref{sec:Discussion} provides a discussion and Section \ref{sec:Conclusion} concludes the paper.

\section{Background and Motivation}
\label{sec:Background}

Blockchain is an append-only distributed ledger replicated across the entire network of \textit{participants} that can record transactions in a verifiable and permanent way using an immutable cryptographic signature.
Participants called \textit{users} propose transactions, 
then send them out to the blockchain network to be verified.
Once a certain amount of proposed transactions have been received,  participants called \textit{block proposers}\footnote{Depending on the technology, these participants are called either miners (Bitcoin), validators (Tendermint), bakers (Tezos) or orderers (Hyperledger Fabric). We choose to generalize them as block proposers as proposed in \cite{Roussille2022}}. attempt to validate them
as an ordered list in a block, i.e., proposes a new block to the system using a consensus protocol (for a recent review on consensus protocols, see \cite{Kaur2021}).

Smart contracts, first suggested in the late 1990s\footnote{ “Smart contracts: Formalizing and securing relationships on
public networks,” First Monday, vol. 2, no. 9, September 1997. [Online]. Available: http://firstmonday.org/ojs/index.php/fm/article/view/548/469}, allow parties to clearly define a contract that is secured by cryptography.
Concretely, a smart contract is a deterministic program implemented using a smart contract language.
Smart contracts are stored as executable bytecodes on the blockchain.
Consequently, they inherit properties like immutability and global distributability. 
Smart contracts are primarily utilized for two objectives:

\begin{itemize}
    \item Keeping funds and state information on the blockchain at the contract's address.
    \item Executing the code or logic to use the funds or modify the contract's state.
\end{itemize}

In order to achieve this, the code is duplicated across multiple nodes of a blockchain, thus taking advantage of the security, permanence, and unchangeability that a blockchain provides. 
This replication also means that when a new block is appended to the blockchain, the same code is essentially executed.

If a transaction has been initiated that meets the specified parameters, the code will proceed with the step associated with those parameters. 
Otherwise, the code will not move forward.
In other words, the code gets executed only due to a transaction call.
A contract can call another one, but a transaction must call the first contract.

Verifying, validating, and testing (VV\&T) of smart contracts became necessary to overcome the extensive range of attacks that may occur due to vulnerabilities \cite{Yamashita2019,Mense2018,Huang2019}.
Benabbou et al. \cite{Benabbou2021} categorizes the available solutions for (VV\&T) of smart contracts into four groups: public test networks, security analysis tools, emulators with smart contract support, and simulators with smart contract support.
They are either specific to some target blockchains, or are specific to some smart contract programming languages, or are limited to a number of vulnerabilities, or are limited to some certain testing levels, or have limited degree of control over the blockchain and smart contract parameters.
Consequently, they conclude that simulators are more appropriate for verifying, validating, and testing smart contracts from all aspects. 
They also provide complete control of the environment for mimicking functioning, enabling us to study the performance closely.

\section{Smart Contract Execution Architectures}
\label{sec:Smart-Contract-Execution-Architectures}

A transaction is an operation that requires changing the status of the blockchain, a message sent from one account to another account that includes a \textit{payload} (binary data) and an amount of crypto currencies.
Each time transactions are created, they follow specific steps depending on the blockchain architecture.
Concretely, there are two main blockchain architectures in the literature.

\subsection{The Order-Execute Architecture}
\label{sec:Execute-After-Architecture}

\begin{figure}
    \centering
    \includegraphics[width=0.80\textwidth]{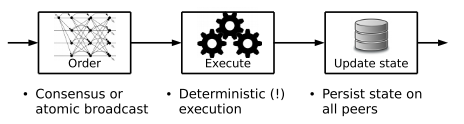}
    \caption{Order-execute architecture \cite{Androulaki2018}}
    \label{fig:Execute-After}
\end{figure}

A majority of the blockchain technologies, permissioned or not (Bitcoin, Ethereum, Tendermint etc.), follow this architecture.
In this architecture, the blockchain participants first 
come to an agreement about the total \textit{order} of transactions,
confirm them as a block, and then the transactions are \textit{executed} by all participants (Figure \ref{fig:Execute-After}).

More precisely, user participants continuously propose transactions to the blockchain system. 
Depending on the consensus algorithm used, one or more \textit{block proposer} participants
gather a block of legitimate transactions (to confirm their validity, they have already been pre-executed).
After a consensus is reached the block is disseminated to the network via a gossip protocol.
Finally, each participant executes the transactions in order (within one block and between blocks) and updates their individual state.
This architecture is illustrated by Figure \ref{fig:Execute-After}.

This architecture is straightforward in its design, making it popular. 
Despite its advantages, this architecture has certain limitations when used in a general-purpose permissioned blockchain.

Since transactions are \textit{executed sequentially}, lower throughput occurs and  this may become a performance bottleneck. 
Due to this, a denial-of-service (DoS) attack can lead to a prolonged execution time, which can significantly reduce throughput. 
To prevent this, certain blockchains such as Ethereum have implemented the concept of gas, which is consumed gradually by each transaction execution.

Moreover, since the transaction execution for validation is not rewarded, rational \textit{block proposer} participants may be incentivized to simply skip it \cite{Luu2015}, potentially sacrificing the overall security and correctness of the system to gain an advantage in both time and computing resources.

Another important issue for this architecture is \textit{non-deterministic} transactions that appear when a general-purpose language, due to its attractiveness and ease of adoption, is used for implementing smart contracts.
Even if the smart contract developer does not introduce non-determinism obviously, unseen details can have the same destructive effect.

Lastly, since all smart contracts are executed on all participants, 
no one can be prevented from accessing the logic of a smart contract, the data of a transaction, or the status of a blockchain.
In other words, executions cannot be done in a \textit{confidential} manner.
Some solutions include advanced zero-knowledge proofs \cite{Sasson2014} and verifiable computation \cite{Kosba2016}, but they are not practical.
    
\subsection{The Execute-Order-Validate Architecture}
\label{sec:The-Execute-Order-Validate-Architecture}

\begin{figure}
    \centering
    \includegraphics[width=0.80\textwidth]{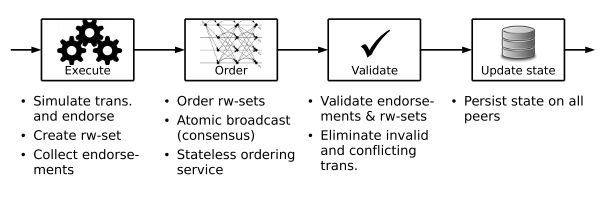}
    \caption{Execute-order-validate architecture \cite{Androulaki2018}}
    \label{fig:Execute-Order}
\end{figure}

This architecture has been proposed in Hyperledger Fabric \cite{Androulaki2018,Ruan2020} to support parallel transactions and improve the throughput of the blockchain. 
In this architecture, the transactions are executed by a designated subset of participants (trusted for this task) before the \textit{block proposer} participants reach a consensus on their order in the blockchain (Figure \ref{fig:Execute-Order}).
After the transactions are ordered in a block, it is enough to diffuse the same state (resulting from the execution) to all participants.
This way, running the same code everywhere is avoided.
     
More precisely, the users first propose transactions consisting of smart contract invocations to the designated subset of participants (not to the whole system, unlike in \textit{order-execute}). 
Then, each one of these participants simulates the execution of the transaction proposals concurrently and returns the results of these simulations to the users. 
The result contains the state update produced by the simulation.
After a user participant receives enough identical results, it bundles them into a single transaction and submits it to the block proposers.

When a block proposer receives transactions, it puts them in a specific order to create a block and then it disseminates the block to the network.
The ordering of transactions is usually by the order a block proposer receives them, however when there are multiple block proposers that receive transactions, the order in a block may not be necessarily the same depending on which block proposer is the one preparing the block (which is dependent on the consensus algorithm used).
What is important here is that the block proposers puts the transactions into a strict order.
Upon receiving a block, all participants sequentially validate the transactions inside using this order.

\section{Use Case: An Industrial Prototype of Trusted Energy Performance Contracts}
\label{sec:Use-Case}

In this section, we present a realistic Trusted Energy Performance Contract (TEPC) use case (previously reported in \cite{Gurcan2018SC}) to evaluate the reliability of the aforementioned smart contract execution architectures.

\subsection{Description}

Energy Performance Contracts (EPCs) \cite{Karakosta2021,Gurcan2018SC} are contracts that compute the energy saving 
to address environmental and economic issues.
The goal here is to carry out an energy audit and guarantee a lasting improvement in the energy efficiency of an existing building or group of buildings belonging to a client (e.g., a business tower, a commercial center, or a municipality). 
It engages a contractor energy service company (usually ESCO) to design and deliver energy measures in order to improve its efficiency by means of reducing energy consumption. 
There can also be third-party stakeholders that provide unbiased measurements if agreed upon.

An EPC system is convenient for being realized by smart contracts so that all stakeholders can trust how the system operates \cite{Gurcan2018SC}.
Hence, to analyse the security vulnerabilities, TEPC is chosen since it involves various types of actors (smart contract users, smart contracts and an oracle) where they interact with each other, and smart contracts rely on the data coming from smart contract users as well as they implement various calculations.
If not designed and developed correctly, such a system may exploit various security vulnerabilities.

\subsection{Model}
\label{sec:Model}

\begin{figure*}[h]
    \centering
    \includegraphics[scale = 0.50]{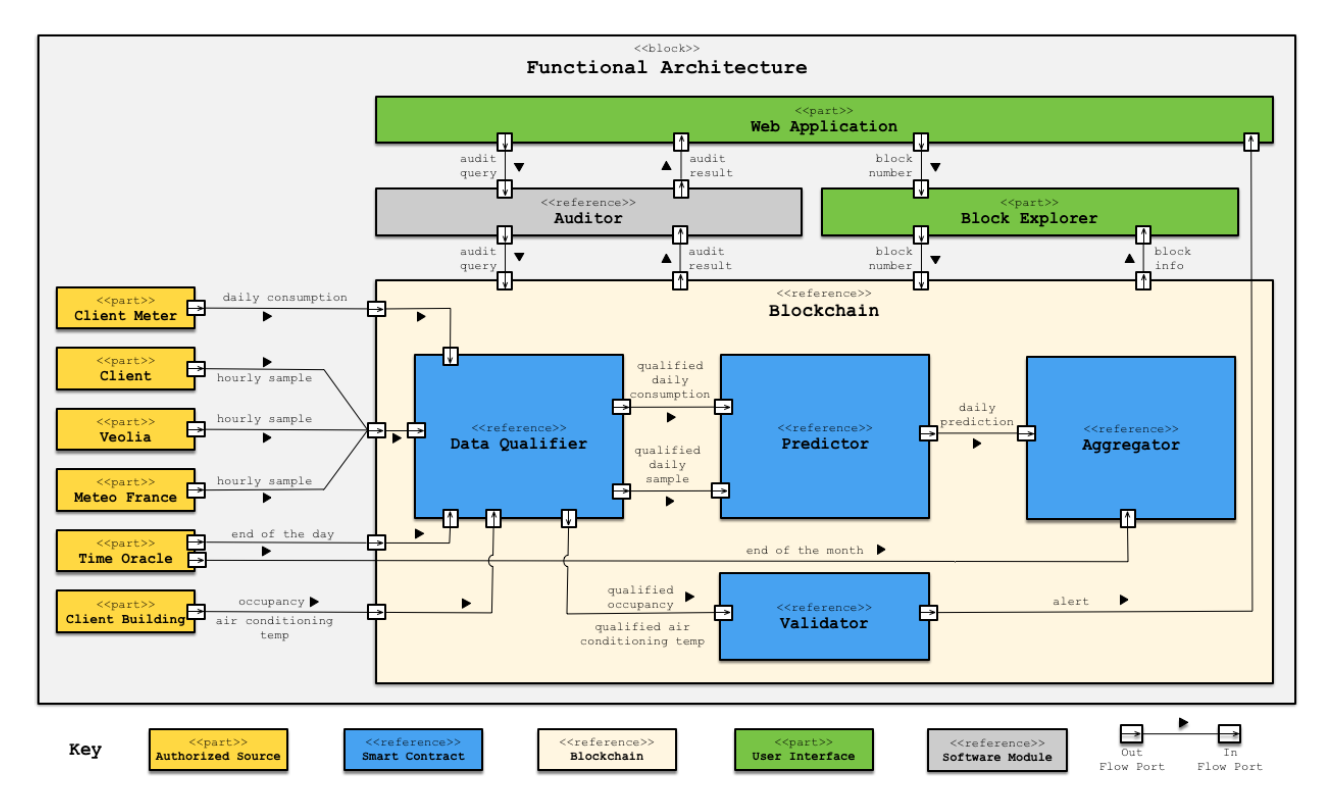}
    \caption{The diagram for the TEPC functional architecture \cite{Gurcan2018SC}.}
    \label{fig:TEPC-Functional-Architecute}
\end{figure*}

We consider the TEPC architectural model presented in \cite{Gurcan2018SC} which is an IoT system deployed on a blockchain (see Figure \ref{fig:TEPC-Functional-Architecute}).

In this model, due to the implementation limitations, the TEPC system is composed of four interconnected smart contracts (\DataQualifier{}, \Predictor{}, \Validator{}, and \Aggregator{}) located on the blockchain and four TEPC authorized sources (\Veolia{}, \MeteoFrance{}, \Client{}, and \ClientMeter{}) located at the intended building and surroundings (the Client Building source is not considered). 
Time Oracle is the one responsible for launching actions at the end of the day or the end of the month.
%
%
To describe precisely the actions done by authorized sources and smart contracts, pseudo codes are given respectively in Algorithm \ref{alg:Pseudo-code-of-EPC-users} and Algorithm \ref{alg:Pseudo-code-of-TEPC-contracts}.

\begin{algorithm}[h!]
\caption{Pseudo code for the TEPC authorized sources.}
\label{alg:Pseudo-code-of-EPC-users}
\scriptsize{
\begin{algorithmic}[1]
\STATE \textbf{\underline{Authorized Stakeholder $u$}}
    \INDONE
    \INDONE \action{at each hour $t$}
        \INDTWO $\tau_{u}^{t}$ $\leftarrow$ senseTemparature()
        \INDTWO $\psi_{u}^{t}$ $\leftarrow$ sensePressure()
        \INDTWO $\rho_{u}^{t}$ $\leftarrow$ senseHumidity()
        \INDTWO $\sigma_{u}^{t} \leftarrow \tau_{u}^{t}, \psi_u, \rho_u$
        \INDTWO proposeTransaction("DataQualifier", "addHourlySample", $\sigma_{u}^{t}$)
        \INDTWO
        
\STATE \textbf{\underline{Authorized Client Meter $m$}}
    \INDONE
    \INDONE \action{at each day at 23:59}
        \INDTWO $\epsilon_{m}^{t} \leftarrow$ senseEnergyConsumption()
        \INDTWO proposeTransaction("DataQualifier", "addDailySample", $\epsilon_{m}^{t}$)
        \INDTWO 
        
\STATE \textbf{\underline{Authorized Time Oracle $o$}}
    \INDONE
    \INDONE \action{at each day at 00:00}
        \INDTWO  proposeTransaction("DataQualifier", "midnightProcess")
        \INDTWO
    \INDONE \action{at the end of the month}
      \INDTWO  proposeTransaction("DataQualifier", "validateMonthlySaving")
      \INDTWO
        
\end{algorithmic}
}
\end{algorithm}

\begin{algorithm*}[h!]
\caption{Pseudo code for the Trusted Energy Performance Contracts.}
\label{alg:Pseudo-code-of-TEPC-contracts}
\scriptsize{
\begin{multicols}{2}
\begin{algorithmic}[1]
\STATE \textbf{\underline{Data Qualifier Contract}}
    \INDONE $\Pi_{u_0}$ : ESCO Samples
    \INDONE $\Pi_{u_1}$ : Meteo France Samples 
    \INDONE $\Pi_{u_2}$ : Client France Samples
    \INDONE $\Pi_{u_3}$ : Client Meter Samples
    \INDONE $\omega_{u_0}$ : Veolia weight
    \INDONE $\omega_{u_1}$ : Meteo France weight 
    \INDONE $\omega_{u_2}$ : Client France weight
    \INDONE $\Gamma$ : Qualified Samples
    \INDONE $p$ : predictor contract
    \INDONE $a$ : aggregator contract
    \INDONE $v$ : validator contract
    \INDONE

\STATE \action{addHourlySample($\sigma_{u}^{t}$)}
    \INDONE $\sigma_{u}^{t} \leftarrow $ $v$.checkVariables($\sigma_{u}^{t}$)
    \INDONE $\Pi_{u} \leftarrow \Pi_{u} \bigcup \{\sigma_{u}^{t}\}$
    \INDONE 
    
\STATE \action{addDailySample($\epsilon_{u}^{t}$)}
    \INDONE $\Pi_{u_3} \leftarrow \Pi_{u_3} \bigcup \{\epsilon_{u}^{t}\}$
    \INDONE 
    
\STATE \action{midnightProcess()}
    \INDONE calculateQualifiedDailySamples()
    \INDONE $p$.predictDailyCons($\Pi_{u_0}$, $\Pi_{u_1}$, $\Pi_{u_2}$, $\Pi_{u_3}$)
    \INDONE $\Pi_{u_0} \leftarrow \emptyset{}$, $\Pi_{u_1} \leftarrow \emptyset{}$, $\Pi_{u_2} \leftarrow \emptyset{}$, $\Pi_{u_3} \leftarrow \emptyset{}$
    \INDONE
    
\STATE \action{monthEndProcess()}
    \INDONE $s$ = $p$.computeMonthlySaving()
    \INDONE $v$.validateMonthlySaving($s$)
    \INDONE $a$.addMonthlySaving($s$)
    \INDONE 
    
\STATE \action{calculateQualifiedDailySamples()}
    \INDONE \textbf{for} ($i \in [0,24]$)
        \INDTWO $\sigma_{u_0}^{t}[h] \leftarrow \Pi_{u_0}[h]$
        \INDTWO $\sigma_{u_1}^{t}[h] \leftarrow \Pi_{u_1}[h]$
        \INDTWO $\sigma_{u_2}^{t}[h] \leftarrow \Pi_{u_2}[h]$
        \INDTWO qualifyByVoting($\sigma_{u_0}^{t}[h], \sigma_{u_1}^{t}[h], \sigma_{u_2}^{t}[h]$)
    \INDONE \textbf{end for}
    \INDONE

\STATE \action{qualifyByVoting($\sigma_{u_0}^{t}[h]$, $\sigma_{u_1}^{t}[h]$, $\sigma_{u_2}^{t}[h]$)}
        \INDONE $g1, g2 \leftarrow $ groupWrtCloseTolerance($\sigma_{u_0}^{t}[h]$, $\sigma_{u_1}^{t}[h]$, $\sigma_{u_2}^{t}[h]$)
        \INDONE $g \leftarrow$ selectGroupWithGreatestWeight($g1,g2$) 
        \INDONE $\Gamma[h] \leftarrow $ averageValue(g)
        \INDONE

\STATE \textbf{\underline{Predictor Contract}}
    \INDONE $a$ : aggregator contract
    \INDONE

\STATE \action{predictDailyCons($\Pi_{u_0}$, $\Pi_{u_1}$, $\Pi_{u_2}$, $\Pi_{u_3}$)}
    \INDONE $\zeta$ = predictCons($\Pi_{u_0}$, $\Pi_{u_1}$, $\Pi_{u_2}$, $\Pi_{u_3}$)
    \INDONE $a$.addDailySaving($\zeta$)
    \INDONE 
    
\STATE \action{predictCons($\Pi_{u_0}$, $\Pi_{u_1}$, $\Pi_{u_2}$, $\Pi_{u_3}$)}
    \INDONE a statistical model provided by the ESCO
    \INDONE
    
\STATE \action{computeMonthlySaving()}
    \INDONE $\Sigma$ = $a$.getDailySavings()
    \INDONE $S$ = $\Sigma$.sum()
    \INDONE

\STATE \textbf{\underline{Aggregator Contract}}
        \INDONE $\Sigma_d$ : list to store daily consumption
        \INDONE $\Sigma_m$ : list to store monthly savings
        \INDONE
\STATE \action{addDailyConsumption($\zeta$)}
    \INDONE $\Sigma_d \leftarrow \Sigma_d \bigcup \{\zeta\}$
    \INDONE 
\STATE \action{addMonthlySaving($s$)}
    \INDONE $\Sigma_m \leftarrow \Sigma_m \bigcup \{s\}$
    \INDONE

\STATE \textbf{\underline{Validator Contract}}
        \INDONE
        
\STATE \action{checkVariable($\sigma_{u}^{t}$)}
        \INDONE $\tau_{u}^{t}$,$\psi_u$, $\rho_u \leftarrow \sigma_{u}^{t}$
        \INDONE \textbf{if} ($\tau_{u}^{t} \leq $ -30) \textbf{then}
            \INDTWO $\tau_{u}^{t}$ = -30
        \INDONE \textbf{else if} ($\tau_{u}^{t} \geq $ 60) \textbf{then}
            \INDTWO $\tau_{u}^{t}$ = 60
        \INDONE \textbf{end if}
        \INDONE
        \INDONE \textbf{if} ($\psi_u^{t} \leq $ 850) \textbf{then}
            \INDTWO $\psi_u^{t}$ = 850
        \INDONE \textbf{else if} ($\psi_u^{t} \geq $ 1060) \textbf{then}
            \INDTWO $\psi_u^{t}$ = 1060
        \INDONE \textbf{end if}
        \INDONE
        \INDONE \textbf{if} ($\rho_u^{t} \leq$ <) \textbf{then}
            \INDTWO $\rho_u^{t}$ = 0
        \INDONE \textbf{else if} ($\rho_u^{t} \geq$  100) \textbf{then}
            \INDTWO $\rho_u^{t}$ = 100
        \INDONE \textbf{end if}
        \INDONE
        
\STATE \action{validateMonthlySaving($ S $)}
        \INDONE \textbf{If} ($S \geq 0$) \textbf{then}
            \INDTWO true
        \INDONE \textbf{else}
            \INDTWO false
        \INDONE \textbf{end if}
        \INDONE
\end{algorithmic} 
\end{multicols}
}
\end{algorithm*}

Each authorized stakeholder $u$ represents a sensor that measures hourly the temperature, humidity, and pressure values (i.e., hourly samples) and proposes a transaction to the \DataQualifier{} smart contract at each hour $t$, containing the dedicated hourly sample $\sigma_{u}^{t}$ of sensed measurements temperature $\tau_{u}^{t}$, pressure $\psi_u^{t}$ and humidity $\rho_u^{t}$, using the action of the form \Function{proposeTransaction()}.  
Similarly, \ClientMeter{} $cm$ is an authorized source representing the electric meter installed at client's premises.
$cm$ regularly measures the amount of electric energy consumed over a time interval using the kilowatt hour (kWh) unit.
It submits the actual daily energy consumption to the \DataQualifier{} smart contract only once at the end of each day.

Lastly, the authorized \TimeOracle{} $o$ provides precise time information such as the end of the day and the end of the month by invoking the  \Function{midnightProcess()} and  \Function{validateMonthlySaving()} methods of the \DataQualifier{} smart contract respectively.

The \DataQualifier{} smart contract has the responsibility of collecting the hourly data samples  $\sigma_{u}^{t}$ coming from authorized stakeholders $u$ by its  \Function{addHourlyDataSample()} method (Algorithm \ref{alg:Pseudo-code-of-TEPC-contracts} line 14).
In that sense, it first calls the  \Function{checkVariables()} of the \Validator{} smart contract (Algorithm \ref{alg:Pseudo-code-of-TEPC-contracts} line 65) to verify whether the data samples are inside the predefined ranges (temperature values between $-30$\degree{}  and  $60$\degree{}, humidity values between $0$\% and $100$\%, pressure values between $850bar$ and $1060bar$ and consumption values higher than or equal to $0kW$) and to replace the violating value in the given sample with the closest boundary value (e.g., a temperature value of -35\degree{} is replaced by -30\degree{}).
After, \DataQualifier{} stores the verified hourly data samples to locally inside a dedicate storage $\Pi_{u}$.
Besides, every midnight, when \TimeOracle{} triggers  \Function{midnightProcess()}, inside the  \Function{calculateQualifiedDailySamples()} method \DataQualifier{} calculates the qualified daily samples based on a voting algorithm that choses the values to use based on a degree of \textit{reliability} calculation:

\begin{itemize}
    \item Group the authorized sources into subgroups of agreement between them with respect to close tolerance.
    \item Select the subgroup with the highest weight.
    \item Compute the qualified value as the average of the source(s) with the greatest weight within the subgroup.
    \item Compute the reliability as the sum of the votes of the chosen group.
\end{itemize}

After that, \DataQualifier{} asks the \Predictor{} smart contract $p$ to predict the daily consumption.
Upon this reuqest, $p$ calculates the predicted consumption based on a statistical model provided by \Veolia{} and then ask \Aggregator{} to store the predicted values as the daily saving by calling its \Function{addDailySaving()} method.  
It also perform a validation of the monthly saving by invoking the \Validator{} smart contract $v$.


The \Predictor{} smart contract $p$ perform only one action of predicting the daily consumption considering $\zeta$ the users $u$ samples ($\Pi_{u_0}$, $\Pi_{u_1}$, $\Pi_{u_2}$) and the \ClientMeter{} $cm$ consumption $\Pi_{u_3}$ and then stock it by invoking the \textit{Aggregator} smart contract $a$.
    
The \Aggregator{} smart contract $a$, on the other hand, is simply responsible for storing the predicted daily consumption values $\zeta$ and monthly savings in a list $\Sigma$, which are later used for billing purposes.

\subsection{Vulnerabilities}
\label{sec:Vulnerabilities}

We exploit some smart contract vulnerabilities considering the three testing levels (unit, integration, and system \cite{Benabbou2021}) to cover various levels. 
To this end, we have chosen the mishandled exceptions, the reentrancy, and the transaction order dependence vulnerabilities, respectively \cite{Mense2018,Tsankov2018,Yamashita2019}.

\textit{Mishandled Exceptions} is an individual smart contract level (i.e., unit level) vulnerability that is tackled by caching the method calls exceptions \cite{Mense2018}. 
There are many situations when an exception can be raised in a smart contract.
An exception can be raised when a client or another smart contract interacts with a smart contract (which is triggered again by a client).
As mentioned in Section \ref{sec:Background}, clients interact with smart contracts by proposing transactions to the blockchain; hence, transaction execution is valid only if no exception is raised.
Besides, exceptions make smart contracts vulnerable to attacks because users will be unaware of any state (or crypto currency) that is lost if these exceptions are not handled properly and the transactions are reverted.
In the context of TEPC, this vulnerability can occur when one or more authorized sources send samples containing \textit{"null"} values, raising an unhandled null pointer exception during the execution.
The process will not be stopped because of the exception raised but will continue without considering these \textit{null} values.

\textit{Reentrancy} is an integration-level vulnerability that occurs when there are conflicting parallel transactions for the same smart contract function \cite{Liu2018,Tsankov2018}.
In the context of TEPC, this vulnerability can occur when a user (\Veolia{}, for example) sends two consecutive samples (not necessarily identical) for the same hour. 
This means having two unexpectedly repeating transactions sent at the same time.
We will consider just the first sample/transaction and ignore the second one.

\textit{Transaction Order Dependence} is a system-level vulnerability that occurs when the order of transactions is manipulated thanks to the timestamp thus enabling some transactions to be exacted before others \cite{Yamashita2019,Munir2023}.
In other words, a transaction that should be executed after another transaction is somehow ordered in the block before that transaction. Thus it is executed before instead of after, causing a different state in the blockchain.
In the context of TEPC, this vulnerability can occur when the midnight process is launched before \ClientMeter{} sends its consumption sample.
In this case, the daily saving of that day will be miscalculated, the execution of the transaction containing \ClientMeter{}'s consumption sample will become unused, and the process for the following day will start.

\section{Simulations}
\label{sec:Simulations}

To evaluate the possible security vulnerabilities concerning TEPC mentioned in Section \ref{sec:Vulnerabilities}, we developed the model given in Section \ref{sec:Model} 
using the MAGE platform, a multi-agent simulation framework for organisation-centric agent-based models\footnote{Multi-AGent Development and Experimentation Platform. (2024). MAGE Platform. Accessed: May 15, 2024. [Online]. Available: \url{https://mage-platform. netlify.app/}}.
We then designed and conducted a variety of simulation experiments.
We consider two cases for all scenarios: (1) the blockchain uses the order-execute (OE) architecture, but for being more scalable, block proposers skip transaction validation, and (2) the blockchain uses the execute-order-validate (EOV) architecture.

Firstly, we designed a success scenario (Section \ref{sec:Scenario01-Success-Scenario}) that provides a baseline validation for the implementation (Section \ref{sec:Use-Case}).
Secondly, we designed deviated scenarios to analyse the vulnerabilities against the smart contract execution architectures.
Four scenarios were set up: single and multiple mishandled handled exceptions, reentrancy, and transaction order dependence.
Thirdly, we systematically explore the impact of vulnerabilities on the successful functioning of the different smart contract execution architectures.

\begin{figure*}[ht]
    \centering
    \subfloat[]{
        \includegraphics[width=0.45\textwidth]{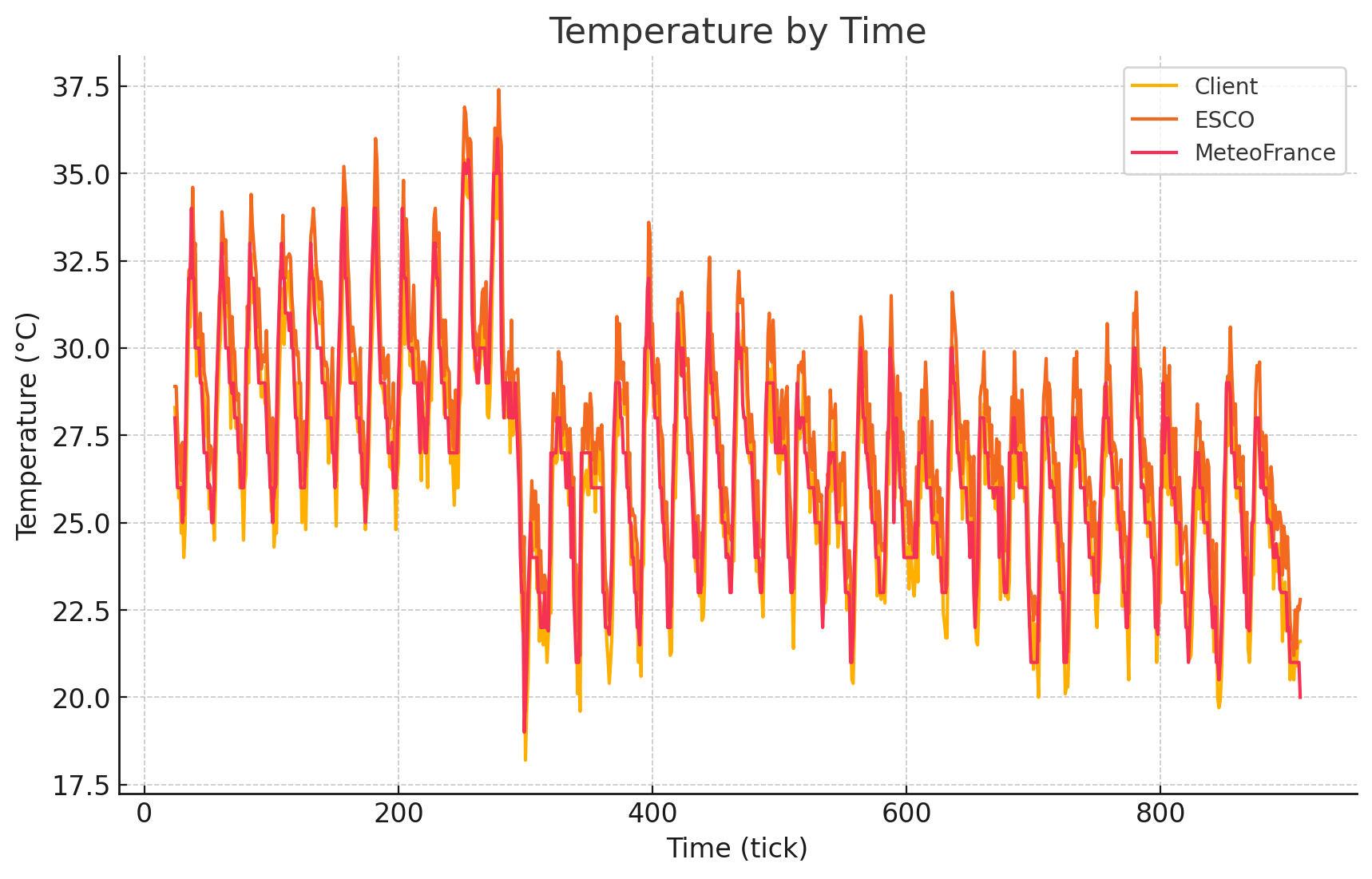}
        \label{fig:Temperature-by-Time}
    }
    \hfill
    \subfloat[]{
        \includegraphics[width=0.45\textwidth]{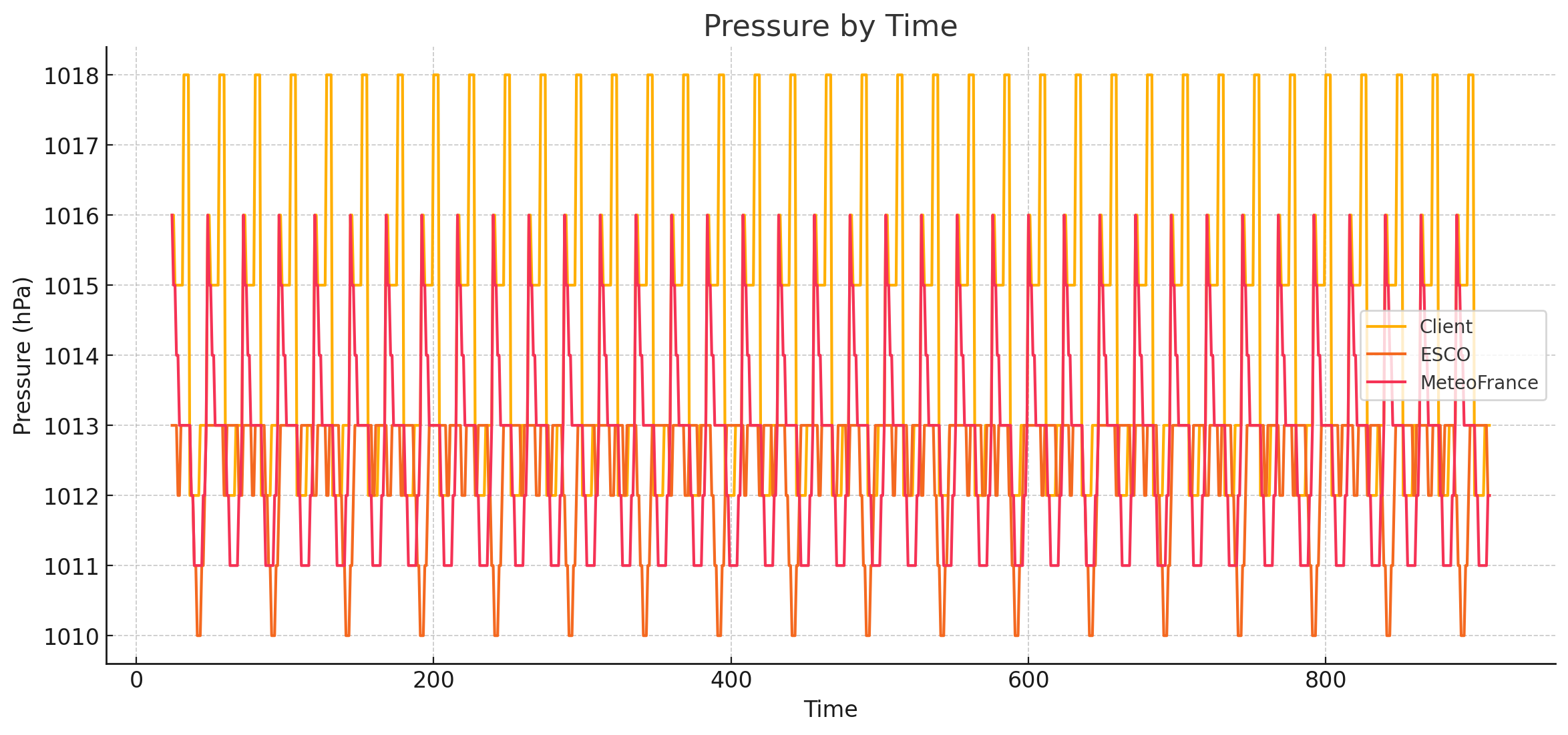}
        \label{fig:Pressure-by-Time}
    }
    \vskip\baselineskip
    \subfloat[]{
        \includegraphics[width=0.45\textwidth]{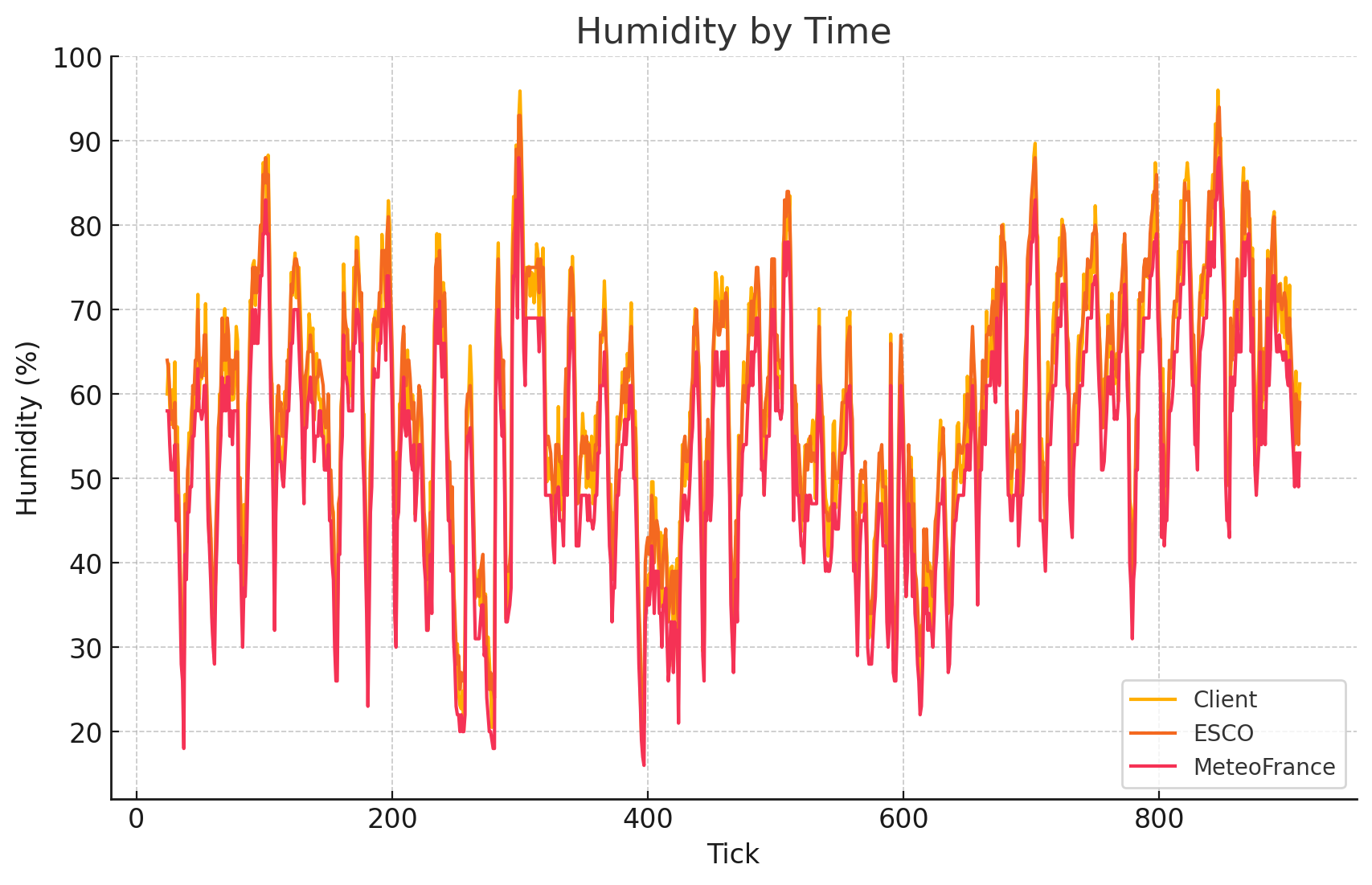}
        \label{fig:Humidity-by-Time}
    }
    \hfill
    \subfloat[]{
        \includegraphics[width=0.45\textwidth]{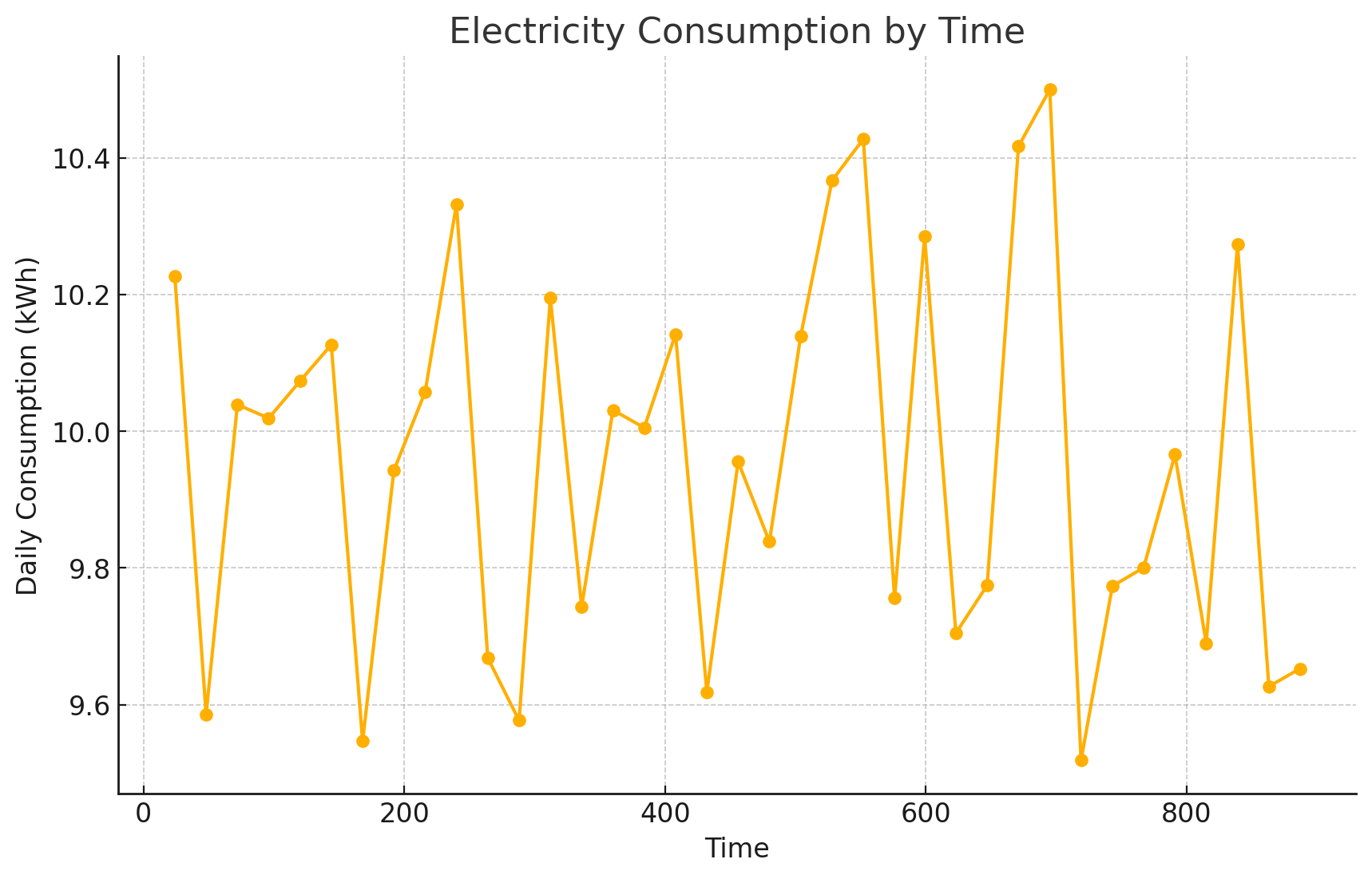}
        \label{fig:Daily-Consumption-by-Time}
    }
    \caption{The illustration of Scenario 1 (baseline) values of Dataset 1. The diagrams (a), (b) and (c) illustrate the hourly values over a month, recorded from three different sources (Client, ESCO, MeteoFrance and Client Meter), while the diagram (d) illustrate the daily consumption measured by ClientMeter.}
    \label{fig:Temperature-by-time}
\end{figure*}

To prepare all these scenarios, we prepared datasets that identify all the samples the authorised sources submit. These are CSV files where, in each row, the datetime and the corresponding sample (i.e., the temperature, the pressure, and the humidity numerical values) are in order.
Simply, by making some sample values \texttt{null}, we can create mishandled exception situations; by repeating a row consecutively, we can create reentrancy situations; and, by advancing the datetime in a row of the dataset of \ClientMeter{}, we can create transaction order dependence situations.

\subsection{Scenario 1: Success Scenario}
\label{sec:Scenario01-Success-Scenario}

\begin{figure*}[ht]
    \centering
    \includegraphics[width=0.99\textwidth]{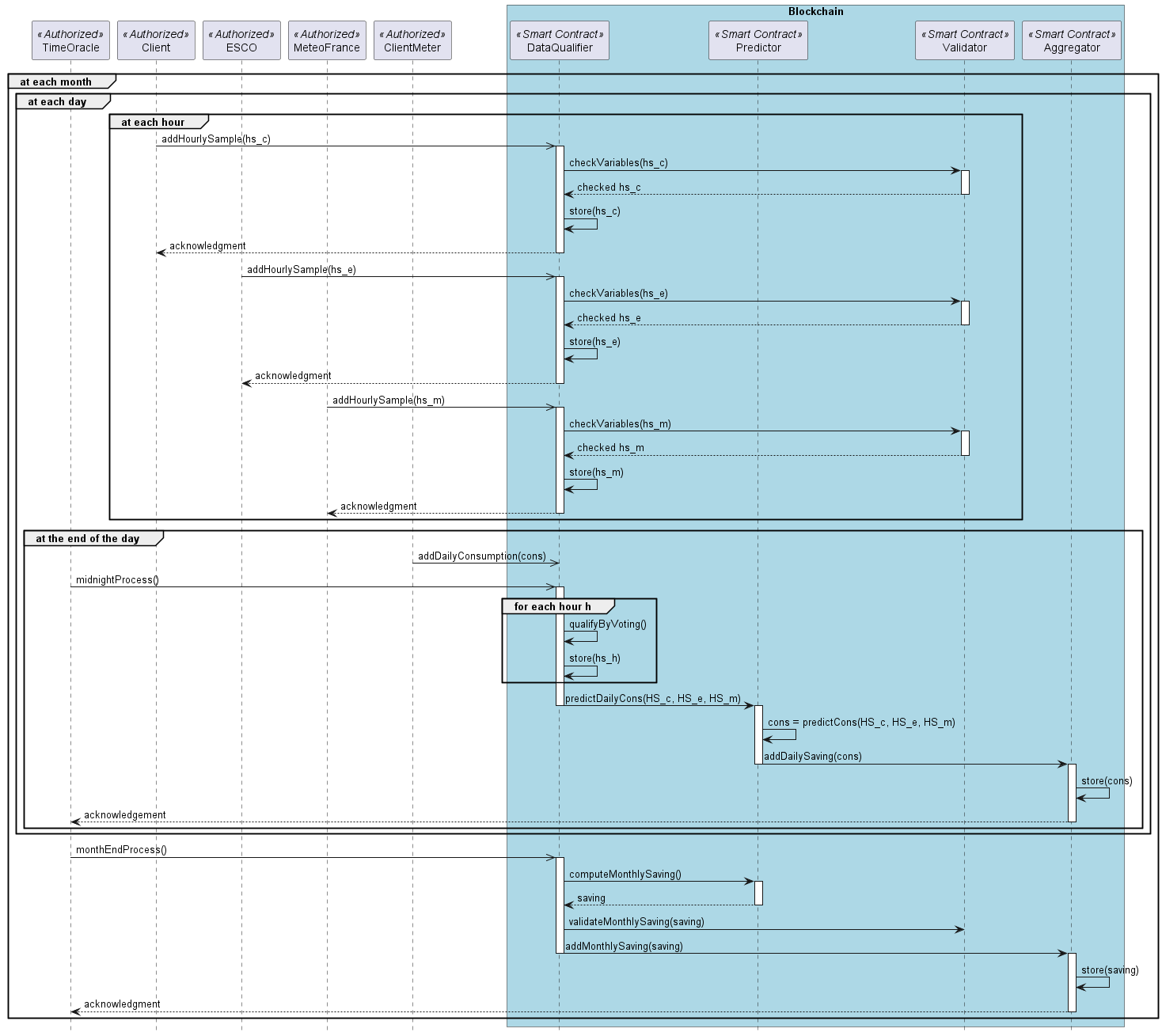}
    \caption{The UML Sequence Diagram for the Overall Monthly TEPC Success Scenario}
    \label{fig:The-Daily-Prediction-Scenario}
\end{figure*}

This scenario shows how the TEPC system is expected to work (see Figure \ref{fig:The-Daily-Prediction-Scenario}).
All samples are supposed to be correctly submitted to the smart contracts (i.e., on time with correct values) and are stored in the blockchain. 
The basis of this scenario is as follows:

\vspace{5pt}

\noindent \textit{On day 1,} \TimeOracle{} deploys the TEPC smart contracts; the \DataQualifier{} will launch the deployment of the \Predictor{}, \Aggregator{}, and \Validator{} contracts.
Then, the authorized sources are initialized using their corresponding datasets.
    
\vspace{5pt}
    
\noindent \textit{From day 2}, starting at 00:00, and until the end of the month, each Authorized Source sends hourly the sensitive samples (temperature, pressure and humidity) to the \DataQualifier{} contract.
Before adding the samples, the \DataQualifier{} calls the \Validator{} contract to check the variables.

\vspace{5pt}    
   
\noindent \textit{From day 2}, starting at 00:00, and until the end of the month, the \ClientMeter{} sends at the end of each day (23:00) the daily energy consumption to the \DataQualifier{} contract.
    
\vspace{5pt}
    
\noindent \textit{From day 2}, at the end of each day (00:00), and until the end of the month,  \TimeOracle{} launches the midnight process by calling the \DataQualifier{} contract.
During this process, the \DataQualifier{} calculates the qualified daily sample considering a voting algorithm.
Then, considering the qualified temperature, pressure, humidity, pressure values, and the energy consumption value sent by the \ClientMeter{}, it predicts the daily consumption by calling the \Predictor{} contract. 
%
After, the \Predictor{} stores the computed consumption in the \Aggregator{} contract.

\vspace{5pt}
    
\noindent \textit{At the end of the month}, \TimeOracle{} invokes the \DataQualifier{} contract to predict the monthly savings made.
The \Predictor{} is called to launch the monthly validation process for this.
In this validation process, the \Predictor{} recovers all the predicted values during the whole month from the \Aggregator{}, computes the saving made and calls the \Validator{} to check if a saving was made or not.

\subsection{Scenario 2a: Single Mishandled Exception}

Here, we consider the occurrence of a single unhandled exception.
We suppose that, at 02:00 of day $2$, due to malfunctioning, the \Client{} source sends a \textit{null} sample, which causes a \textit{null} pointer exception to the \Validator{} smart contract while checking the variables in the sample. 

\subsection{Scenario 2b: Multiple Mishandled Exceptions}

Here, we consider the occurrence of several unhandled exceptions.
It is supposed that, during the whole month, the \Client{} source arbitrarily sends \textit{null} samples (in total between 25 to 30 times), which causes unhandled null pointer exceptions in the  \Validator{} smart contract. 

\subsection{Scenario 3: Reentrancy}

Here, we suppose that the \Veolia{} source arbitrarily sends two different samples for the same hour during the whole month (in total between 25 to 30 times): 
correct temperature, pressure and humidity samples at first and immediately after (at the same hour) \textit{0} values for all due to malfunctioning.

\subsection{Scenario 4: Transaction Order Dependence}

Here, we suppose that the \textit{Client Meter} source's energy consumption transactions sometimes have a transmission delay\footnote{Transmission delay is the time taken for a packet to be transmitted from one node to another over the network. The transmission delay depends on factors like the distance between the nodes, the medium's speed, and the packet's size. Such delays can occur for various reasons, such as network congestion, packet loss, and processing overheads when a protocol architecture like CoAP together with IPv6 is used. Reducing transmission delays is essential to ensure efficient communication and timely response. This can be achieved using congestion control, packet prioritization, and efficient routing algorithms. However, this is not in the scope of this article.} and they sometimes arrive on time and sometimes even after midnight (on days 3, 11, 16, 25, 27, and 30).

\section{Results}
\label{sec:Results}

Each experiment is initialized using the parameter values from Table \ref{tab:Simulation-Parameters} with four different datasets. Additionally, every experiment is repeated once for the Order-Execute architecture and once for the Execute-Order-Validate architecture.

\begin{table}[h]
\centering
\caption{Simulation Parameters}
\label{tab:Simulation-Parameters}
\begin{tabular}{|l|c|}
\hline
\textbf{Parameter}        & \multicolumn{1}{l|}{\textbf{Value}} \\ \hline
Simulation step           & 1                                   \\ \hline
Block creation frequency  & 1                                   \\ \hline
Day length in hours       & 24                                  \\ \hline
Month length in hours     & 24*31                               \\ \hline
Deployment of contracts   & 12                                  \\ \hline
Initialization of sources & 16                                  \\ \hline
Month beginning (1st day) & 24                                  \\ \hline
Midnight beginning        & 48                                  \\ \hline
End of month              & 24*31                               \\ \hline
\end{tabular}
\end{table}

Briefly, each simulation step represents an hour and each day the authorized sources submit data at each hour, which makes the day length 24 hours.
Simultaneously, the blockchain system generates blocks every hour, grouping together the data for the same hour within the same block.
For the purposes of the simulation, we are assuming that each month consists of 31 days, with the first day being reserved for setting up the TEPC system, including deploying the smart contracts and configuring the authorized sources.


\begin{table}[h!]
\centering
\caption{Results obtained upon the execution of all scenarios over four different datasets.}
\begin{tabular}{|l|c|c|}
\hline
Scenario & OE & EOV \\
\hline
\multicolumn{3}{|c|}{Dataset1} \\
\hline
Scenario 1 & 8149.419 & 8149.419 \\
Scenario 2a & 8142.096 & 8142.078 \\
Scenario 2b & 8088.869 & 8087.419 \\
Scenario 3 & 8146.246 & 8149.419 \\
Scenario 4 & 8036.419 & 8036.419 \\
\hline
\multicolumn{3}{|c|}{Dataset2} \\
\hline
Scenario 1 & 8304.302 & 8304.302 \\
Scenario 2a & 8302.392 & 8302.375 \\
Scenario 2b & 8295.392 & 8295.122 \\
Scenario 3 & 8301.547 & 8304.302 \\
Scenario 4 & 8201.547 & 8201.547 \\
\hline
\multicolumn{3}{|c|}{Dataset3} \\
\hline
Scenario 1 & 10301.437 & 10301.437 \\
Scenario 2a & 10299.947 & 10299.931 \\
Scenario 2b & 10293.141 & 10292.852 \\
Scenario 3 & 10298.348 & 10301.437 \\
Scenario 4 & 10217.537 & 10217.537 \\
\hline
\multicolumn{3}{|c|}{Dataset4} \\
\hline
Scenario 1 & 10213.697 & 10213.697 \\
Scenario 2a & 10211.364 & 10211.348 \\
Scenario 2b & 10207.076 & 10206.076 \\
Scenario 3 & 102013.530 & 10213.697 \\
Scenario 4 & 10129.797 & 10129.797 \\
\hline
\end{tabular}
\label{tab:All-Results}
\end{table}

All obtained results are collected in Table \ref{tab:All-Results}.
For each scenario, for each smart contract execution architecture and each dataset, the resulting monthly saving values are plotted.
As expected, for Scenario 1, both experiments always have the same monthly saving value calculated.

Concerning Scenarios 2a and 2b, exceptions are raised upon the \textit{"null"} value.
In the Execute-Order-Validate (EOV) architecture, the system excludes the faulty transactions from the next block, and the samples from the other authorized sources are typically processed.
However, in the Order-Execute (OE) architecture, when exceptions occur, the execution of the whole stops since a faulty transaction is enough to invalidate a whole block.
Note that the monthly saving values have deviated for both architectures, but for the OE ones, the deviations are more significant.

Concerning Scenario 3, while in the EOV architecture, the second fault transaction is omitted and not included in the block, in the OE architecture, the second transaction overrides the stored sample value (with zeros) since also because there is no additional control implemented inside the smart contract against such a case.

Lastly, concerning Scenario 4, both execution architectures behave the same way, including the delayed transaction to a later block. Because of this, the monthly savings are miscalculated in both of them.
Since this situation is independent of the execution architecture, to avoid this problem, before calculating the daily savings, the smart developer needs to ensure that the consumption value from the \ClientMeter{} source is received.

\section{Discussion}
\label{sec:Discussion}

This paper delves into the critical importance of ensuring software reliability \cite{Sanchez2018} in blockchain-based IoT systems through security analysis and smart contract execution. 
We provide an overview of the current state of research, highlight some of the critical challenges that need to be addressed, and perform a simulation-based evaluation of the two smart contract execution architectures on a realistic blockchain-based IoT energy performance contract case study. 
As pointed out by \cite{Paulavicius2021}, blockchain simulators are crucial for understanding these complex systems. However, as far as we know, no simulator is specifically created to assess various blockchain architectures. Most simulators concentrate on individual blockchain systems and primarily evaluate performance rather than security aspects.

\begin{table}[h!]
    \centering
    \caption{Comparison of reliability on different levels between OE and EOV architectures}
    \resizebox{\textwidth}{!}{
    \begin{tabular}{|>{\raggedright}m{4cm}|c|c|}
        \hline
        \textbf{Reliability Aspect} & \textbf{Order-Execute (OE)} & \textbf{Execute-Order-Validate (EOV)} \\
        \hline
        Smart Contract (Unit-) Level& Lower & Higher \\
        \hline
        Application (Integration-) Level & Higher & Lower \\
        \hline
        Blockchain System (System-) Level & Less Efficient & More Efficient \\
        \hline
        Overall Reliability & Moderate & High \\
        \hline
    \end{tabular}
    }
    \label{tab:comparison}
\end{table}

The results show that at the unit level, the Execute-Order-Validate (EOV) architecture demonstrates superior reliability in smart contract execution compared to the Order-Execute (OE) approach, since EOV will exclude transactions that are causing errors during execution.
This enhancement at the unit level is critical because it ensures that individual smart contracts are executed with a higher degree of correctness and fewer errors. 

Integration-level reliability, however, reveals a trade-off. While the EOV architecture excels at the unit level, it faces challenges in application-level integration, where the OE architecture shows higher reliability. 
The results show that, the OE architecture is generally better in this level because it allows for better control and understanding of transaction flow before execution. 
By ordering transactions before execution, the system ensures that all dependencies and interactions are well understood, reducing the risks. 
The OE architecture provides predictable execution, maintains atomicity and consistency, and facilitates easier auditing and analysis of transactions, enhancing security. In contrast, the EOV architecture, where transactions are executed immediately and validated post-execution, faces challenges with unpredictable states and complex validation, making it more susceptible to such vulnerabilities.

At the system-level, on the other hand, the EOV architecture is more promising than the OE architecture in terms of reliability.
This distinction underscores the imperative of fault prevention strategies in designing and implementing smart contracts, where the EOV architecture exemplifies a proactive approach to mitigating risks before they manifest into systemic failures.
To the best of our knowledge, no other study in the literature has developed such an evaluation which is based on the three different levels of vulnerabilities.

\section{Conclusion}
\label{sec:Conclusion}

We conclude that security analysis for reliable smart contract execution in IoT systems is a critical area of software reliability and dependability research, particularly in the context of safety-critical and economically significant systems, that needs to be addressed to ensure these systems' reliability and security. 

By understanding the security requirements of the blockchain-based system, the security analysis process can be tailored to meet the system's specific needs.
With the proper security measures in place, blockchain-based IoT Systems can be used to execute smart contracts securely and provide a secure and reliable platform for businesses and users.

In conclusion, our study offers novel insights into the challenges and methodologies for securing smart contract execution in blockchain-based IoT systems, paving the way for future research that further explores and refines these approaches. 
Our work underscores the importance of integrating fault prevention, fault tolerance, and reliability prediction strategies from the early stages of system development to deployment, ensuring blockchain-based IoT systems' safety, security, and dependability in a rapidly evolving digital landscape.



 \bibliographystyle{elsarticle-num} 
 \bibliography{cas-refs}

\begin{thebibliography}{10}
\expandafter\ifx\csname url\endcsname\relax
  \def\url#1{\texttt{#1}}\fi
\expandafter\ifx\csname urlprefix\endcsname\relax\def\urlprefix{URL }\fi
\expandafter\ifx\csname href\endcsname\relax
  \def\href#1#2{#2} \def\path#1{#1}\fi

\bibitem{Abdelhamid2019}
M.~Abdelhamid, G.~Hassan, Blockchain and smart contracts, 2019, p. 91 – 95.
\newblock \href {https://doi.org/10.1145/3328833.3328857}
  {\path{doi:10.1145/3328833.3328857}}.

\bibitem{Sharma2023}
P.~Sharma, R.~Jindal, M.~D. Borah, A review of smart contract-based platforms,
  applications, and challenges, Cluster Computing 26~(1) (2023) 395 – 421.
\newblock \href {https://doi.org/10.1007/s10586-021-03491-1}
  {\path{doi:10.1007/s10586-021-03491-1}}.

\bibitem{Suvitha2021}
M.~Suvitha, R.~Subha, A survey on smart contract platforms and features, 2021,
  p. 1536 – 1539.
\newblock \href {https://doi.org/10.1109/ICACCS51430.2021.9441970}
  {\path{doi:10.1109/ICACCS51430.2021.9441970}}.

\bibitem{Sharma2021}
J.~Sharma, J.~Dubey, A study of blockchain for secure smart contract, Lecture
  Notes in Networks and Systems 204 LNNS (2021) 591 – 598.
\newblock \href {https://doi.org/10.1007/978-981-16-1395-1\_45}
  {\path{doi:10.1007/978-981-16-1395-1\_45}}.

\bibitem{Wood2014}
G.~Wood, Ethereum: A secure decentralised generalised transaction ledger
  (2014).

\bibitem{Yulianto2023}
S.~Yulianto, H.~L. Hendric Spits~Warnars, H.~Prabowo, Meyliana, A.~N.
  Hidayanto, Security risks and best practices for blockchain and smart
  contracts: A systematic literature review, 2023, p. 282 – 287.
\newblock \href {https://doi.org/10.1109/ICIMTech59029.2023.10278055}
  {\path{doi:10.1109/ICIMTech59029.2023.10278055}}.

\bibitem{Roussille2023}
H.~Roussille, O.~G\"{u}rcan, F.~Michel, A taxonomy of blockchain incentive
  vulnerabilities for networked intelligent systems, IEEE Communications
  Magazine 61~(8) (2023) 108--114.
\newblock \href {https://doi.org/10.1109/MCOM.005.2200904}
  {\path{doi:10.1109/MCOM.005.2200904}}.

\bibitem{Benabbou2021}
C.~Benabbou, O.~G\"{u}rcan, A survey of verification, validation and testing
  solutions for smart contracts, in: 2021 Third International Conference on
  Blockchain Computing and Applications (BCCA), 2021, pp. 57--64.
\newblock \href {https://doi.org/10.1109/BCCA53669.2021.9657040}
  {\path{doi:10.1109/BCCA53669.2021.9657040}}.

\bibitem{Ruan2020}
P.~Ruan, D.~Loghin, Q.-T. Ta, M.~Zhang, G.~Chen, B.~C. Ooi, A transactional
  perspective on execute-order-validate blockchains, in: Proceedings of the
  2020 ACM SIGMOD International Conference on Management of Data, SIGMOD '20,
  Association for Computing Machinery, New York, NY, USA, 2020, p. 543–557.
\newblock \href {https://doi.org/10.1145/3318464.3389693}
  {\path{doi:10.1145/3318464.3389693}}.

\bibitem{Androulaki2018}
E.~Androulaki, A.~Barger, V.~Bortnikov, C.~Cachin, K.~Christidis, A.~D. Caro,
  D.~Enyeart, C.~Ferris, G.~Laventman, Y.~Manevich, S.~Muralidharan, C.~Murthy,
  B.~Nguyen, M.~Sethi, G.~Singh, K.~Smith, A.~Sorniotti, C.~Stathakopoulou,
  M.~Vukolic, S.~Cocco, J.~Yellick, Hyperledger fabric: a distributed operating
  system for permissioned blockchains, Proc of the 13th EuroSys Conf (2018).

\bibitem{Gurcan2018SC}
Önder Gürcan, M.~Agenis-Nevers, Y.~Batany, M.~Elmtiri, F.~L. Fevre,
  S.~Tucci-Piergiovanni, \textit{"An Industrial Prototype of Trusted Energy
  Performance Contracts Using Blockchain Technologies"}, in: 2018 IEEE 20th Int
  Conf on HPC and Comm, IEEE 16th Int Conf on Smart City, IEEE 4th Int Conf on
  Data Sci and Syst, 2018.
\newblock \href {https://doi.org/10.1109/HPCC/SmartCity/DSS.2018.00222}
  {\path{doi:10.1109/HPCC/SmartCity/DSS.2018.00222}}.

\bibitem{Roussille2022}
H.~Roussille, O.~G\"{u}rcan, F.~Michel, Agr4bs: A generic multi-agent
  organizational model for blockchain systems, Big Data and Cognitive Computing
  6~(1) (2022) 41p.
\newblock \href {https://doi.org/10.3390/bdcc6010001}
  {\path{doi:10.3390/bdcc6010001}}.

\bibitem{Kaur2021}
M.~Kaur, S.~Gupta, Blockchain consensus protocols: State-of-the-art and future
  directions, in: International Conference on Technological Advancements and
  Innovations (ICTAI), 2021, pp. 446--453.
\newblock \href {https://doi.org/10.1109/ICTAI53825.2021.9673260}
  {\path{doi:10.1109/ICTAI53825.2021.9673260}}.

\bibitem{Yamashita2019}
K.~Yamashita, Y.~Nomura, E.~Zhou, B.~Pi, S.~Jun, \textit{"Potential Risks of
  Hyperledger Fabric Smart Contracts"}, in ieee int workshop on blockchain
  oriented software engineering (iwbose) (2019).

\bibitem{Mense2018}
A.~Mense, M.~Flatscher, \textit{"Security Vulnerabilities in Ethereum Smart
  Contracts"}, Proceedings of the 20th International Conference on Information
  Integration and Web-based Applications \& Services (2018).

\bibitem{Huang2019}
Y.~Huang, Y.~Bian, R.~Li, J.~L. Zhao, P.~Shi, Smart contract security: A
  software lifecycle perspective, IEEE Access 7 (2019) 150184--150202.
\newblock \href {https://doi.org/10.1109/ACCESS.2019.2946988}
  {\path{doi:10.1109/ACCESS.2019.2946988}}.

\bibitem{Luu2015}
L.~Luu, J.~Teutsch, R.~Kulkarni, P.~Saxena, {Demystifying incentives in the
  consensus computer}, Proc. of the ACM Conference on Computer and
  Communications Security 2015-Octob (2015) 706--719.
\newblock \href {https://doi.org/10.1145/2810103.2813659}
  {\path{doi:10.1145/2810103.2813659}}.

\bibitem{Sasson2014}
E.~Ben~Sasson, A.~Chiesa, C.~Garman, M.~Green, I.~Miers, E.~Tromer, M.~Virza,
  Zerocash: Decentralized anonymous payments from bitcoin, in: 2014 IEEE
  Symposium on Security and Privacy, 2014, pp. 459--474.
\newblock \href {https://doi.org/10.1109/SP.2014.36}
  {\path{doi:10.1109/SP.2014.36}}.

\bibitem{Kosba2016}
A.~Kosba, A.~Miller, E.~Shi, Z.~Wen, C.~Papamanthou, Hawk: The blockchain model
  of cryptography and privacy-preserving smart contracts, in: 2016 IEEE
  Symposium on Security and Privacy (SP), 2016, pp. 839--858.
\newblock \href {https://doi.org/10.1109/SP.2016.55}
  {\path{doi:10.1109/SP.2016.55}}.

\bibitem{Karakosta2021}
C.~Karakosta, A.~Papapostolou, G.~Vasileiou, J.~Psarras, 3 - financial schemes
  for energy efficiency projects: lessons learnt from in-country
  demonstrations, in: D.~Borge-Diez, E.~Rosales-Asensio (Eds.), Energy Services
  Fundamentals and Financing, Energy Services and Management, Academic Press,
  2021, pp. 55--78.
\newblock \href
  {https://doi.org/https://doi.org/10.1016/B978-0-12-820592-1.00003-8}
  {\path{doi:https://doi.org/10.1016/B978-0-12-820592-1.00003-8}}.

\bibitem{Tsankov2018}
P.~Tsankov, A.~M. Dan, D.~Drachsler-Cohen, A.~Gervais, F.~Buenzli, M.~T.
  Vechev, \textit{"Securify: Practical Security Analysis of Smart Contracts"},
  Proc of the 2018 ACM SIGSAC Conference on Computer and Communications
  Security (2018).

\bibitem{Liu2018}
C.~Liu, H.~Liu, Z.~Cao, Z.~Chen, B.~Chen, A.~W. Roscoe, \textit{"ReGuard:
  Finding Reentrancy Bugs in Smart Contracts"}, 2018 IEEE/ACM 40th
  International Conference on Software Engineering: Companion (ICSE-Companion)
  (2018) 65--68.

\bibitem{Munir2023}
S.~Munir, C.~Reichenbach, Todler: A transaction ordering dependency analyzer -
  for ethereum smart contracts, in: 2023 IEEE/ACM 6th International Workshop on
  Emerging Trends in Software Engineering for Blockchain (WETSEB), 2023, pp.
  9--16.
\newblock \href {https://doi.org/10.1109/WETSEB59161.2023.00007}
  {\path{doi:10.1109/WETSEB59161.2023.00007}}.

\bibitem{Sanchez2018}
C.~Sánchez, G.~Schneider, M.~Leucker, Reliable smart contracts:
  State-of-the-art, applications, challenges and future directions, Lecture
  Notes in Computer Science (including subseries Lecture Notes in Artificial
  Intelligence and Lecture Notes in Bioinformatics) 11247 LNCS (2018) 275 –
  279.
\newblock \href {https://doi.org/10.1007/978-3-030-03427-6\_21}
  {\path{doi:10.1007/978-3-030-03427-6\_21}}.

\bibitem{Paulavicius2021}
R.~Paulavičius, S.~Grigaitis, E.~Filatovas, A systematic review and empirical
  analysis of blockchain simulators, IEEE Access 9 (2021) 38010--38028.
\newblock \href {https://doi.org/10.1109/ACCESS.2021.3063324}
  {\path{doi:10.1109/ACCESS.2021.3063324}}.

\end{thebibliography}





\end{document}